\documentclass[prl,aps,amssymb,amsmath,superscriptaddress,twocolumn,showpacs]{revtex4}

\newcommand{\const}{\mathrm{const}\ }

\renewcommand{\epsilon}{\varepsilon}

\newcommand{\kk}{\mathbf{k}}

\newcommand{\p}{\mathbf{p}}
\renewcommand{\phi}{\varphi}
\newcommand{\R}{\mathbb{R}}

\newcommand{\x}{\mathbf{x}}

\DeclareMathOperator{\Tr}{Tr}

\begin{document}

\title{Bi-Polaron and $N$-Polaron Binding Energies}
\thanks{\copyright\, 2010 by the authors. This paper may be reproduced, in its entirety, for non-commercial purposes.}

\author{Rupert L. Frank}
\email{rlfrank@math.princeton.edu}
\affiliation{Department of Mathematics,
Princeton University, Washington Road, Princeton, NJ 08544, USA}

\author{Elliott H. Lieb}
\email{lieb@princeton.edu}
\affiliation{Department of Mathematics,
Princeton University, Washington Road, Princeton, NJ 08544, USA}
\affiliation{Department of Physics,
Princeton University, P.~O.~Box 708, Princeton, NJ 08542, USA}

\author{Robert Seiringer}
\email{rseiring@princeton.edu}
\affiliation{Department of Physics,
Princeton University, P.~O.~Box 708, Princeton, NJ 08542, USA}

\author{Lawrence E. Thomas}
\email{let@virginia.edu}
\affiliation{Department of Mathematics, University of Virginia, Charlottesville, VA 22904, USA}

\begin{abstract}
  The binding of polarons, or its absence, is an old and subtle topic.
  Here we prove two things rigorously. First, the transition from
  many-body collapse to the existence of a thermodynamic limit for $N$
  polarons occurs precisely at $U=2\alpha$, where $U$ is the
  electronic Coulomb repulsion and $\alpha$ is the polaron coupling
  constant. Second, if $U$ is large enough, there is no multi-polaron
  binding of any kind.  Considering the known fact that there \emph{is}
  binding for some $U>2\alpha$, these conclusions are not
  obvious and their proof has been an open problem for some time.
\end{abstract}


\maketitle

The (large) polaron, first considered by H. Fr\"ohlich \cite{Fr} in
1937, is a model of an electron moving in three dimensions and interacting
with the quantized optical modes of a polar crystal. In suitable
units, its Hamiltonian is
\begin{align}\label{eq:h1}
 H^{(1)} =p^2 & + \int a^\dagger(\kk)a(\kk)\,d\kk \\ 
& + \frac{\sqrt{\alpha}}{\sqrt 2\, \pi} \int \frac{1}{k}[a(\kk)\exp(i\kk\cdot\x) +h.c.] \,d\kk \,, \notag
\end{align}
where $a(\kk)$ are the annihilation operators  of the scalar, longitudinal modes (with $[a(\kk),a^\dagger(\kk')]=\delta(\kk-\kk')$), $\p$ is the
momentum of an electron, and $\alpha $ is the
coupling constant. (Other authors have used a different convention, where $\alpha$ is replaced by $\alpha/\sqrt2$ \cite{Fr,DePeVe}.) In the ground state, with energy $E^{(1)}$, the electron accompanied by the localized
excitations of the phonon field constitutes the polaron. Through the years, the polaron
has served both as a model  for an electron in an ionic crystal
and as
a simple model for a dressed particle in nonrelativistic quantum field theory.

Of great physical interest is the binding energy of $N$ polarons, with Hamiltonian
\begin{align}
 H^{(N)}_U = & \sum_{j=1}^N p_{j}^2 + \int a^\dagger(\kk)a(\kk)\,d\kk \\
& + \frac{\sqrt{\alpha}}{\sqrt2\,\pi} \sum_{j=1}^N\int \frac{1}{k}[a(\kk)\exp(i\kk\cdot\x_j)+h.c.]\,d\kk \notag \\
& +U \sum_{1\leq i<j\leq N} |\x_i -\x_j|^{-1} \,, \notag
\end{align}
and ground state energy $E^{(N)}_U$. Here, $U\geq 0$ is the Coulomb repulsion parameter, equal to $e^2$. The derivation of $H^{(N)}_U$ in \cite{Fr} implies that $U>2\alpha$,  and this is crucial for thermodynamic stability, as we shall see.

We first consider the bipolaron binding energy $\Delta E_U= 2 E^{(1)}-E^{(2)}_U$.
For some time this was thought to be zero for all $U\geq 2\alpha$, on the basis of an 
inadequate variational calculation, but it is now known \cite{DePeVe} to be positive for some $U>2\alpha$. The question we address is whether $\Delta E_U=0$ for $U$ sufficiently large. We will show that there is a finite constant $C$ such that $\Delta E_U=0$ if $U/\alpha\geq C$. 
It is understood that the effective interaction induced by the phonon field for two polarons at large distances $d$ is approximately Coulomb-like $-2\alpha/d$, but this alone does not preclude binding. The known existence of bipolarons for some $U>2\alpha$ is an effect of correlations. It is a priori conceivable that correlations lead to an effective attraction that is  stronger than Coulomb at large distances.  If it were, for example,
equal to $(2\alpha/d)\log(\log(\log(d)))$, then this minuscule perturbation of
Coulomb's law, which would be virtually undetectable by a variational calculation, would result in binding for \emph{all} $U$. The finiteness of $C$ is a problem that has resisted a definitive resolution for many years.

The second problem we consider is the existence of the thermodynamic
limit. For large $N$, physical intuition suggests that $E^{(N)}_U\sim
-\const N$. This supposition is known to be false if $U<2\alpha$.
Indeed, it was shown in \cite{GrMo} that, even with the Pauli
principle, $E^{(N)}_U\sim -\const N^{7/3}$ when $U<2\alpha$. Absent
the Pauli principle, $E^{(N)}_U$ would behave even worse, as $-\const
N^3$. It is also known \cite{GrMo} that $E^{(N)}_U\geq -\const N^{2}$
if $U>2\alpha$.  The latter bound ought to be $-\const N$ instead, and
we prove this for all $U> 2\alpha$. Even more is true; there is a
number $U_c(\alpha)$ such that when $U\geq U_c(\alpha)$, then
$E^{(N)}_U=N E^{(1)}$, i.e., there is no binding whatsoever. There
will, of course, be an intermediate region in which bound complexes
form, a gas of bipolarons, for example, or a crystal. The Pauli
principle plays no role in our considerations, and our results hold
equally for fermions and bosons.

The following rigorous results concerning $E^{(1)}$ will be important
in our analysis.  (i) For all $\alpha$, $E^{(1)}\leq -\alpha$
\cite{LePi}.  For small $\alpha$, $E^{(1)}\sim -\alpha$ according to
the lower bound in \cite{LiYa}, which is $E^{(1)}\geq
-\alpha-\alpha^2/3$. (ii) For all $\alpha$, $E^{(1)}\leq - C_P
\alpha^2$ \cite{Pe}, where $C_P=0.109$ is the number determined by Pekar's
integral equation \cite{Mi}. (iii) Asymptotically, as
$\alpha\to\infty$, $E^{(1)}\sim -C_P\alpha^2$ according to
\cite{DoVa,LiTh}.  (iv) There is a representation for $E^{(N)}_0$ in
terms of path integrals.  In terms of the partition function
$Z^{(N)}(T)=\Tr \exp\big(-T H^{(N)}_0\big)$, $E^{(N)}_0 =
-\lim_{T\to\infty}T^{-1}\log Z^{(N)}(T)$.  (Strictly speaking,
$Z^{(N)}(T)$ does not exist because of the translation invariance of
$H^{(N)}_0$, and the infinite number of phonon modes. These
technicalities can be handled by inserting appropriate cutoffs, to be
removed at the end of the calculation \cite{Ro,Sp}.) It was shown in
\cite{Fe} that after one integrates out the phonon variables,
$Z^{(N)}(T)$ has a functional integral representation
\begin{equation}
 \label{eq:trace}
Z^{(N)}(T) = \int d\mu^{(N)} \exp\left[ \frac\alpha2 \sum_{i,j} \int_0^T\!\! \int_0^T \!\frac{e^{-|t-s|}\,dt\,ds}{|\x_i(t)-\x_j(s)|} \right] \,,
\end{equation}
where $d\mu^{(N)}$ is Wiener measure on all $T$-periodic paths $(\x_1(t),\ldots,\x_N(t))$. (Strictly speaking, $t-s$ has to be understood modulo $T$, but this is irrelevant as $T\to \infty$.)


We shall now state our results as three theorems, and sketch their proofs. While we postpone the discussion of technical details to a subsequent paper, the full structure and concept of the proofs are visible in this letter. We use the symbols $c_1, c_2, \ldots$ to denote various calculable positive constants that arise in the proof.  Our results are strongest in the bipolaron case, which is also easier than the general case and illustrates our concepts most clearly. 


\textbf{Theorem 1 (Absence of bipolaron binding).} \textit{There is a computable constant $C$, such
that if $U>  C\alpha$, then the energy expectation $\langle \Psi, H^{(2)}_U\Psi\rangle$ for any bipolaron
wave function $\Psi$ is strictly bigger than $2E^{(1)}$, i.e., there is no binding if $U > C \alpha$.}

It is convenient to structure the proof in four steps.

\emph{Step 1. Partition of the interparticle distance:} We fix a
length $\ell$, whose value will later be chosen proportional to
$\alpha^{-1}$, and partition the relative distance $r=|\x_1-\x_2|$
between the particles into spherical shell-like regions of radial size
$2^{k-1}\ell\leq r\leq 2^k\ell$ with $k=1,2,\ldots$. This partitioning
is one of the key points of our analysis. In addition there is the
$k=0$ region, where the particle separation is between zero and
$\ell$. Because of the uncertainty principle these regions have to
overlap a bit, but this can be easily handled and we ignore it for for
simplicity. There is a kinetic energy cost for localizing the
particles according to this partition, which is $c_1 2^{-2k} \ell^{-2}$
in the shell $k$. In the next step we look at the energy of the
particles localized to one of these shell-like regions.

\emph{Step 2. Further localization for well-separated particles:} For
$k\geq 1$ we further localize the particles into individual boxes of
size $2^{k-3}\ell$. This costs another localization error $c_2 2^{-2k} \ell^{-2}$. Because the separation exceeds $2^{k-1}\ell$, the two particles cannot be in the same or neighboring boxes.  From the path
integral \eqref{eq:trace}, but now with the $\x_i(t)$'s constrained to
their respective boxes, we see that \emph{the separated particles feel
  an effective Coulomb-like attractive potential}. However, this can
contribute at worst $-c_3\alpha 2^{-k} \ell^{-1}$ to the energy. But
the Coulomb repulsion is at least $U 2^{-k}\ell^{-1}$, which implies
that the total energy exceeds $2E^{(1)}$ if
\begin{equation}
 \label{eq:klarge}
U 2^{-k}\ell^{-1} > c_3\alpha 2^{-k} \ell^{-1} + (c_1+c_2) 2^{-2k} \ell^{-2} \,.
\end{equation}
If this inequality holds for $k=1$, it holds for all $k\geq 2$ as well. Thus, if we 
can deal with the $k=0$ region, we will establish that binding is not possible if 
\begin{equation}
\label{eq:klarge1}
 U\alpha^{-1} > c_3 + (c_1+c_2)/(2\ell\alpha) \,.
\end{equation}

\emph{Step 3. The region of no minimal separation:}
In the $k=0$ region, the Coulomb repulsion is at least $U \ell^{-1}$, but, since there is no minimal separation, we have no direct handle on the possible attraction due to the field. We need a lemma, which we will prove in Step 4. It concerns $E^{(2)}_0$, the energy of the bipolaron with no Coulomb repulsion, i.e., $U=0$;
\begin{equation}
 \label{eq:binding}
E^{(2)}_0 \geq 2E^{(1)} - 7\alpha^2/3
\quad \text{for all}\ \alpha \,.
\end{equation}
Assuming this, the total energy in the $k=0$ region exceeds $2E^{(1)}$ provided
\begin{equation}
 U \ell^{-1} > 7\alpha^2/3 + c_1 \ell^{-2} \,,
\end{equation}
that is, no binding occurs if
\begin{equation}
\label{eq:ksmall1}
 U\alpha^{-1} > 7\ell\alpha/3 + c_1 /(\ell\alpha) \,.
\end{equation}
Setting the right sides of \eqref{eq:klarge1} and \eqref{eq:ksmall1} equal leads to
 the choice $\ell=c_4/\alpha$ and to absence of binding if $U>C\alpha$, as asserted.

\emph{Step 4. The universal lower bound \eqref{eq:binding}:}
In this step, $U=0$ and we denote the $\alpha$-dependence of energies explicitly. We first note that 
\begin{equation}
 \label{eq:twiceenergy}
E^{(1)}(2\alpha)\geq 2 E^{(1)}(\alpha)-4\alpha^2/3 \,.
\end{equation}
This follows from the lower bound $E^{(1)}(\alpha)\geq -\alpha -\alpha^2/3$ in \cite{LiYa} and the upper bound $E^{(1)}(\alpha)\leq -\alpha$ in \cite{LePi}, stated above. So \eqref{eq:binding} will follow if we can prove that
 \begin{equation}
 \label{eq:twoenergy}
E^{(2)}_0(\alpha)\geq E^{(1)}(2\alpha)-\alpha^2 \,.
\end{equation}
For this purpose we go back to the functional integral
\eqref{eq:trace} and use Schwarz's inequality $\langle e^{a+b} \rangle
\leq \langle e^{2a} \rangle^{1/2} \langle e^{2b} \rangle^{1/2}$, where
$\langle\cdot\rangle$ now denotes expectation with respect to Wiener
measure.  We choose $a$ to be the sum of the two terms $i=j=1$ and
$i=j=2$ in \eqref{eq:trace}, and $b$ to be the mixed terms $i\neq j$.
Since $\langle e^{2a} \rangle^{1/2} \sim e^{-T E^{(1)}(2\alpha)}$ for
large $T$, inequality \eqref{eq:twoenergy} will be achieved if we can
show that $\langle e^{2b} \rangle^{1/2}\sim e^{T\alpha^2}$.  At first
sight, the double path integral $\langle e^{2b} \rangle$ looks like
that for a positronium-like atom, i.e., two particles attracting each
other through a Coulomb force with coupling constant $4\alpha$. The
trouble is that the interaction in \eqref{eq:trace} is at different
times, i.e., $|x_1(t)-x_2(s)|^{-1}$. A simple application of Jensen's
inequality, however, shows that we can fix the time difference $u=t-s$
and bound
$$
\langle e^{2b} \rangle
\!\leq \!\!\int_{-\infty}^\infty \!\!\frac{e^{-|u|} du}2 \!\int\! d\mu^{(2)} \!\exp\!\left[ 4\alpha \!\!\int_0^T \!\!\!\frac{\,dt}{|\x_1(t)-\x_2(t-u)|} \right]
$$
Because of the $T$-periodic time translation invariance of the Wiener
measure, the path integral is, in fact, independent of $u$. Hence we
get the positronium-like answer as a bound.  This completes our
argument for the universal bound (\ref{eq:binding}), and hence the absence of bipolaron
binding for sufficiently large $U/\alpha$.



We now return to the case of general particle number $N$.
As we noted above, there is no thermodynamic limit if $U<2\alpha$, even with Fermi 
statistics \cite{GrMo}. In contrast,

\textbf{Theorem 2 (Thermodynamic stability for the $N$-polaron system).} \textit{If $U>2\alpha $ the energy of $N $ particles  (polarons) is bounded below 
as 
$$
E_U^{(N)} \geq -C (U, \alpha) \ N
$$
where $C(U,\alpha)$ is finite and independent of $N$, but can depend on $U, \alpha$ and on statistics.}

Our upper bound on $C(U,\alpha) $ goes to $+\infty$ as $U$ goes down
to $2\alpha$, but we are not certain that this divergence reflects the true
situation.

Theorem 2 implies the existence of the thermodynamic limit, $\lim_{N\to\infty} N^{-1} E_U^{(N)}=C'(U,\alpha)$, when $U>2\alpha$. The reason is that $E^{(N)}_U$ is sub-additive, i.e., $E^{(N+M)}_U\leq E^{(N)}_U +E^{(M)}_U$ (by considering $N$ particles in a ground state located near Princeton and $M$ particles in a ground state located near Charlottesville). This fact, together with the linear lower bound from Theorem 2, implies the existence of the thermodynamic limit, see \cite[Sec. 14.2]{LiSe}. 

Theorem 2 is an essential ingredient for our proof of  Theorem 3 about the absence of any binding for large enough $U$.

The proof of Theorem 2 does not use the partitioning and localization
of Theorem 1. Instead we bound $H^{(N)}_U$ from below by the average
over translations of a Hamiltonian pertaining to  a finite-size box and
with a short range, i.e., Yukawa-like interaction.  This `sliding
method' of localization was introduced in \cite{CoLiYa}, and later
used in \cite{LiSo}, to analyze Foldy's law for bosonic jellium.

The localization is accomplished by choosing a function $\chi(\x)$ with finite range $L$ and integral $\int \chi(\x)^2 d\x=1$. Next, consider the function $f(\x) = (1-e^{-\omega |\x|}\chi*\chi(\x))/|\x|$, where $*$ means convolution. Essentially, $f$ is the difference of the Coulomb potential and a cut-off Yukawa potential. If $\omega$ is large enough, $f$ will be positive definite, i.e., have a positive Fourier transform, as shown in \cite{CoLiYa}. The crucial inequality, then, is
\begin{align}\label{eq:posdef}
\iint d\x\,d\x' & \left( \sum\nolimits_i \delta(\x-\x_i) -\rho^\dagger(\x)\right) f(\x-\x') \notag \\
& \quad \times\left( \sum\nolimits_j \delta(\x'-\x_j) -\rho(\x')\right) \geq 0 
\end{align}
for any $\x$-dependent operator $\rho(\x)$. We apply this to
$$
\rho(\x)= \frac{\sqrt2\,\pi}{\sqrt{\alpha}} \int k\, e^{i\kk\cdot\x} a(\kk) \, d\kk \,.
$$
In
the classical jellium case, $\rho=\rho^\dagger$ is just the background density,
whereas here it is a fluctuating quantum field. 
If we multiply out the various terms in \eqref{eq:posdef} the resulting inequality can be written as
$$
H^{(N)}_U \geq L^{-3} \int_{\R^3} d\mathbf{z} \, H_{\textbf{z}} + (U-2\alpha) \sum_{i<j} |\x_i-\x_j|^{-1} \,,
$$
where $H_{\textbf{z}}$ is a Hamiltonian of the particles that happen to lie in a box centered at $\textbf{z}$ and with sides of length $L$. It is crucial to note the fact that $f(0)$ is finite, which allows us to replace the apparent, unwanted Coulomb self-energy in \eqref{eq:posdef} by a term linear in $N$.

The particles interact through Yukawa-like forces. The field appears in $H_\mathbf{z}$ as $\rho(\x)\chi(\x-\mathbf z)$, so it is localized to the box as well. Thus $H_\mathbf{z}$ refers to a quantum-mechanical problem confined to a box of a fixed size but with an indeterminate number $n_\mathbf z$ of particles.

Once the localization to finite-size boxes is established we can
follow the analysis in \cite{GrMo}, based on the commutator bounds in
\cite{LiTh}, to show that $H_\mathbf z$ is bounded from below by $-
c_5\alpha^2 n_\mathbf z^2 -D(\alpha,L)n_\mathbf z$, where $D(\alpha,L)\to \infty$ as $L\to
0$. On the other hand, 
$$
\sum_{i<j} |\x_i-\x_j|^{-1} \geq c_6 L^{-4}
\int d\mathbf{z}\, n_{\textbf{z}} (n_{\mathbf{z}}-1) \,.
$$ For any
$U>2\alpha$, the length $L$ can thus be chosen small enough such that
$(U-2\alpha)c_6/L > c_5$. The price paid for this is the energy $-D(\alpha,L)
N$ which can be large, but is finite. This concludes the proof of
thermodynamic stability.



As $U$ increases from $2\alpha$ the system is thus stable but can form many-body bound complexes such as the bipolaron -- perhaps even a periodic super-lattice. This is a largely unexplored area. Eventually, no binding is possible, as the following theorem asserts.

\textbf{Theorem 3 (No binding for large $U$).} \textit{There is a computable constant $U_c(\alpha)$ such that if $U\geq U_c(\alpha)$, then the ground state energy equals $E_U^{(N)}=N E^{(1)}$ for all $N$.}

Although the optimum value of $U_c(\alpha)$ might depend on particle statistics, our bound does not. We can prove that $U_c(\alpha)\leq \const\alpha$ for large $\alpha$. We believe, but we cannot prove, that this linear law holds for \emph{all} $\alpha$.

\emph{Step 1.} We use a similar partitioning as in Theorem~1, but relative to \emph{nearest neighbor} distances. That is, each $\x_i$ is localized in some shell-like region $2^{k_i-1}\ell \leq t_i \leq 2^{k_i} \ell$, where $t_i$ is the distance between particle $i$ and its nearest neighbor in the configuration $(\x_1,\x_2,\ldots,\x_N)$.

Then, as in Step 2 of Theorem~1, we localize each particle $i$ in a box whose size is smaller than, but comparable to, $2^{k_i} \ell$. As in the bipolaron case we have to remember the kinetic energy associated with this two-fold localization. It is a geometric fact that any $\x_i$ can be the nearest neighbor of at most $12$ other particles. This allows us to compensate the localization energy by sacrificing part of the Coulomb repulsion.

\emph{Step 2.} With every particle thus localized in some box we write the functional integral for the ground state energy as in \eqref{eq:trace}, except that we now include the Coulomb repulsion as well as the polaronic attraction terms. The exponential now contains
$$
\frac\alpha2 \!\sum_{i,j=1}^N \!\int_0^T\!\! \int_0^T \!\frac{e^{-|t-s|}\,dt\,ds}{|\x_i(t)-\x_j(s)|}
-U \!\!\!\!\!\sum_{1\leq i<j\leq N} \!\int_0^T \!\!\!\frac{dt}{|\x_i(t)-\x_j(t)|}
$$
We relabel the particles so that $k_1=\ldots=k_M=0$ and $k_{M+1},\ldots,k_N\geq 1$, i.e., such that particles $1,2,\ldots,M$ are precisely those having a nearest neighbor within distance $\ell$. Accordingly, we split the sum in the exponential into three pieces. The first one corresponds to the total energy of the $M$ particles with $k_i=0$,
$$
\frac\alpha2 \!\sum_{i,j=1}^M \!\int_0^T\!\! \int_0^T \!\frac{e^{-|t-s|}\,dt\,ds}{|\x_i(t)-\x_j(s)|}
-U \!\!\!\!\!\sum_{1\leq i<j\leq M} \!\int_0^T \!\!\!\frac{dt}{|\x_i(t)-\x_j(t)|}
$$
The second piece corresponds to the polaronic self-energy for particles $i>M$, that is,
$$
\frac\alpha2 \sum_{i=M+1}^N \int_0^T\!\! \int_0^T \!\frac{e^{-|t-s|}\,dt\,ds}{|\x_i(t)-\x_i(s)|} \,.
$$
The third is
\begin{align*}
 \sum_{j=M+1}^N \sum_{i=1}^{j-1} & \left( \alpha\int_0^T\!\! \int_0^T \!\frac{e^{-|t-s|}\,dt\,ds}{|\x_i(t)-\x_j(s)|} \right. \\
& \left. \quad -U \int_0^T \frac{dt}{|\x_i(t)-\x_j(t)|} \right) \,.
\end{align*}

If we keep in mind the confinement of the particles to their individual boxes, and how the distances between these boxes are related to their sizes, we see that the third piece is necessarily negative provided $U/\alpha$ is large. This condition is independent of the parameter $\ell$. We are left with the first and second piece. The second just gives us the energy $(N-M) E^{(1)}$ after integration. For the first piece we write $U=U_1+U_2$ with $U_1>2\alpha$. Since the $U_2$-part of the Coulomb repulsion is bounded from below by $\const U_2 M/\ell$ by construction, Theorem~2 shows that the total energy of the first piece is bounded from below by $-C(U_1,\alpha) M + \const U_2 M/\ell$. This energy will be bigger than the sum of $ M E^{(1)}$ and the ($\ell$-dependent) localization error, provided $U_2$ is large enough. This completes the proof of Theorem~3.

The authors are grateful to Herbert Spohn for making us aware of the problem of proving
absence of binding for large $U$. Grants from the U.S.~National Science Foundation are gratefully
acknowledged:  PHY-0652854 (E.L. and R.F.),  PHY-0845292 (R.S.).


\bibliographystyle{amsalpha}

\end{document}